# ÆtherFlow: Principled Wireless Support in SDN

Muxi Yan, Jasson Casey, Prithviraj Shome and Alex Sprintson
Texas A&M University
College Station, Texas, USA
{mxyan, jasson.casey, prithvirajhi, spalex}@tamu.edu

Andrew Sutton
University of Akron
Akron, Ohio, USA
asutton@uakron.edu

*Abstract*—Software Defined Networking (SDN) drastically changes the meaning and process of designing, building, testing, and operating networks. The current support for wireless networking in SDN technologies has lagged behind its development and deployment for wired networks. The purpose of this work is to bring principled support for wireless access networks so that they can receive the same level of programmability as wireline interfaces. Specifically we aim to integrate wireless protocols into the general SDN framework by proposing a new set of abstractions in wireless devices and the interfaces to manipulate them. We validate our approach by implementing our design as an extension of an existing OpenFlow data plane and deploying it in an IEEE 802.11 access point. We demonstrate the viability of software-defined wireless access networks by developing and testing a wireless handoff application. The results of the experiment show that our framework is capable of providing new capabilities in an efficient manner.

## I. Introduction

Software Defined Networking (SDN) has recently emerged as a transformational tool to design and operate communication networks and services. The core principle of the SDN paradigm is a separation of the network control and data planes. It enables network administrators to have a centralized view of the network and provides a standardized interface for remote configuration of network devices. In particular, the SDN approach provides an abstraction of the underlying data plane and an interface to manipulate that abstraction. This approach provides the capability to manage and operate a large network through a logically centralized controller and to define custom network behaviors.

While the SDN approach has significant benefits for both wireline and wireless networks, the academic and industrial communities have focused primarily on wireline networks, while wireless networks have received significantly less attention. Currently published SDN standards, the most popular of which is OpenFlow [6], do not provide support for wireless protocols, which poses a major obstacle to developing SDN-enabled heterogeneous networks with wireless components. Attempts to support wireless networking within that framework have been ad hoc, and true network visibility is missing with respect to wireless protocols.

The goal of this work is to fill the gap by extending the basic concepts of SDN to support wireless networks in a principled

This material is based upon work supported by the National Science Foundation under Grants No. 1422655, 1423322 and by the AFOSR under contract No. FA9550-13-1-0008.

way. Note that any reasonable design must not be specific to a single protocol or implementation of SDN, but applicable to every viable implementation. Furthermore, any solution must not be tailored to a single application, but enable potentially *any* application. Some examples are:

- *Wireless handoff* efficiently manages the Layer 2 transition of a client between APs (access points).
- *Client steering* optimizes the association of all the mobile stations in a wireless network area by directing a client to connect to specific APs based on signal strength and current usage.
- *Mobile station and user-based QoS control* implements QoS policies specifically for wireless users and their applications (e.g., throttling high-traffic applications for guest users of a network).

To support a broad range of applications, our approach is to extend a generalized model of SDN derived from the OpenFlow specifications [2]. These extensions include support for wireless ports and channels as well as the events and counters specific to wireless networks and devices. Our model enables SDN controllers to configure, query, and control IEEE 802.11 Access Points (APs), and allows them to respond to a wide range of wireless events.

To validate the approach, we have implemented the model as an extension of the OpenFlow protocol, with a corresponding software implementation in the CPqD SoftSwitch software data plane [1]. We refer to our extension of this model and our initial implementation as *ÆtherFlow*. We tested the extension by developing a wireless mobility application that supports Layer 2 handoff of mobile stations between IEEE 802.11 access points. The resulting system performs on par with a traditional switching method. Moreover, this shows that ÆtherFlow provides a solid foundation for more intelligent wireless SDN applications, which is our long-term goal for this research.

Our contributions can be summarized as follows.

- The extension of the generic SDN model to provide explicit support for wireless radio interfaces and wireless access points.
- An implementation of this extension based on the OpenFlow protocol and the CPqD SoftSwitch.
- An implementation of a controller application using ÆtherFlow framework and experiments to demonstrate the viability of SDN-controlled access points for efficient wireless handoff.

## II. RELATED WORK

The interest of extending WLAN capabilities has been a community goal for a long time, but traditional methods have certain constraints. For example, the approaches reported in [7], [9] require modifications to the mobile clients (referred to as *mobile stations* in the Wi-Fi standard), which makes those approaches hard to deploy and test.

A recent technical report by the Open Networking Foundation (ONF) [4] identified the challenges of mobile networks, such as scalability, management, flexibility and cost, and provided a brief discussion of how SDN solutions can address these issues in few specific scenarios. A working group of Open Networking Foundation, Wireless & Mobile Working Group (WMWG), has been focusing on devising new SDN architecture for wireless use cases of different types [3]. However, to the best of our knowledge no concrete solutions were proposed by either ONF or WMWG up to now.

Several previous works presented systems that use OpenFlow extensions to achieve specific goals in wireless networks. In particular, OpenRoad [11], [13], [12] proposes to use the OpenFlow framework as a research platform for Wi-Fi and Wi-MAX systems. The platform supports slicing and virtualization of network resources, allowing different experimental services to run at the same time. SoftCell [5] focused on LTE networks and proposed to integrate SDN framework into the LTE core network architecture. The objective of ÆtherFlow is to design data plane control interfaces for wireless ports, which is different from these projects.

Other attempts to apply SDN to IEEE 802.11 networks include Odin [10] and OpenSDWN [8]. They provide certain wireless interface control and configuration capabilities to the SDN controller. In these solutions, virtual access points and associated device contexts are created for each individual mobile device, and move across access points when the client handoff occurs. Such type of framework can handle user mobility gracefully, but results in overhead in terms of both computational load and traffic load during handoff, especially in the settings with a large number of clients and high user mobility. ÆtherFlow offers a set of interfaces that costs less but still supports a wide variety of wireless applications.

In contrast to the existing works, ÆtherFlow provides a principled and general definition of wireless abstractions within an existing SDN framework. Our approach only requires incremental modifications to the existing SDN network elements.

## III. OPENFLOW EXTENSIONS

In our previous work, we derived a generalized SDN abstractions model, called TinyNBI [2], from the OpenFlow specifications [6]. In TinyNBI, the OpenFlow data plane is composed of several elements. The data plane elements and their structural relationships are depicted as UML diagram in Figure 1. TinyNBI model provides a clean low-level interpretation of the core OpenFlow abstractions and supports development of higher layer abstractions through refinement

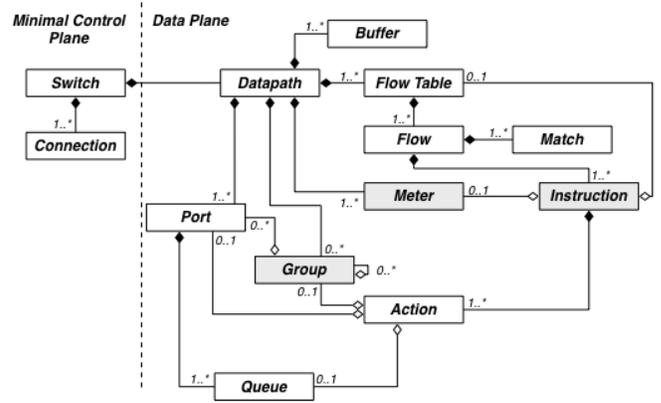

Fig. 1: A UML diagram of OpenFlow data model. Each box represents a data plane element and the lines show the dependencies relationship among the elements.

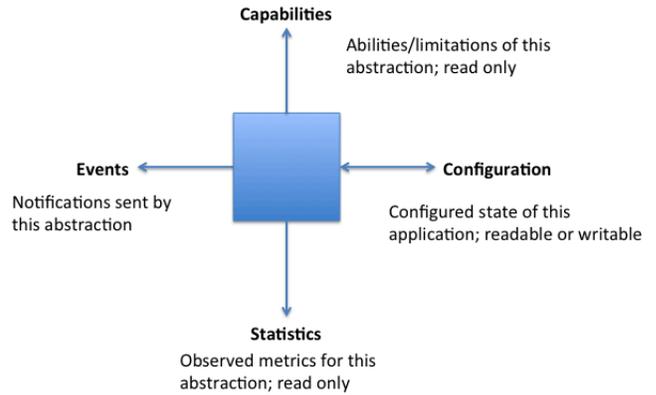

Fig. 2: OpenFlow interfaces abstraction.

or extension. The model is primarily based on the notion of resources shared across a data plane.

In the TinyNBI model, each component exposes four types of interfaces: capabilities, configuration, statistics and events. These interfaces and their information flow directions are conceptually depicted in Figure 2.

**Capabilities.** Not every switch provides the same level of support for e.g., matching on protocol fields. Each component in the model must provide a facility that allows a controller to discover the set of operations supported by the device.

**Configuration.** Each OpenFlow data model element has some configurable parameters that can be modified during switch operation. The configuration interfaces are used to modify state and behavior of the data model element.

**Statistics.** OpenFlow switches gather statistical information via counters such as the numbers of bytes received and transferred over a port. Statistics interfaces provide the controller with access to the current state and values of these counters.

**Events.** The events interface reports to the controller certain types of events that occur during switch operation.

## A. Wireless Data Plane Abstractions

In order to support wireless networks, we refine the notion of ports from the TinyNBI SDN model. The SDN model already has a distinction between physical and logical ports. A *physical port* corresponds to an actual interface (e.g., Ethernet card), whereas a *logical port* is typically defined by software. Logical ports are often used for protocol tunneling and link aggregation.

To support wireless SDN controllers we introduce new types of both physical and logical ports. ÆtherFlow introduces wireless physical port corresponding to an IEEE 802.11 (commonly known as WiFi) radio interface. This allows controllers to query and configure the physical device over which packets are sent and received.

Because a single 802.11 radio interface can support multiple simultaneous wireless access points (APs), ÆtherFlow also introduces wireless logical port. Each wireless logical port is associated with its underlying physical port.

For packet processing, whenever a packet from a wireless AP is processed, its metadata records its *input port* as the logical port for the AP and its *input physical port* as the physical port the AP is created on. The frames received on the wireless interface are adapted into regular Ethernet frames for pipeline processing, meaning that we do not have to define any new protocol matching features for 802.11 MAC frame fields. This also allows an existing SDN implementation to compose wireless logical ports into link aggregation ports or various forms of tunnels.

The new data plane elements (wireless ports) defined in ÆtherFlow expose to the controller a set of control interfaces, categorized in the same way as the TinyNBI model, which are described as below.

**Capabilities.** ÆtherFlow allows the controller to query and obtain the capabilities of the radio interfaces of an AP. The supported capabilities information for wireless physical port includes (i) IEEE 802.11 version; (ii) channels; (iii) transmission power; (iv) encryption and key management methods; (v) maximum number of APs supported. ÆtherFlow does not define capabilities interface for wireless logical port.

**Configuration.** An OpenFlow controller can use ÆtherFlow messages to create or remove AP and dynamically (re)configure the following properties of an AP:

- Wireless physical port: (i) IEEE 802.11 version; (ii) channel; (iii) transmission power.
- Wireless logical port: (i) SSID; (ii) BSSID; (iii) encryption and key management method.

In addition, ÆtherFlow allows the controller to change the state of mobile stations associated with it, e.g., drop a station. Any new configuration to an AP is immediately applied. The configuration interfaces provide a high degree of programmability to applications that require these parameters to be adjusted during network operation.

**Events.** An SDN controller can receive MAC layer events related to a mobile station. ÆtherFlow currently supports the following types of events for wireless logical port: (i) probe; (ii) authentication; (iii) deauthentication; (iv) association; (v) reassociation; (vi) disassociation; (vii) authorization. ÆtherFlow does not define any event interface for wireless physical port.

The events occur when AP receives the corresponding 802.11 management frames. With these events reported, the controller can keep track of the 802.11 state of all the mobile stations communicating with the APs under control.

**Statistics.** An SDN controller can query the statistics of each physical wireless port and its associated logical ports. For a wireless logical port the following types of statistics are supported: (i) number of packets sent and received; (ii) number of bytes sent and received; (iii) number of retries; (iv) number of retry failures; (v) current signal strength of a station; (vi) average signal strength of a station; (vii) connection duration of a station. For wireless physical port the set of supported statistics is identical to that supported by the OpenFlow protocol.

## B. Messages

To implement ÆtherFlow in the framework of OpenFlow, we use experimenter messages provided in OpenFlow protocol to carry ÆtherFlow messages. In the current version, nine messages are defined in ÆtherFlow:

- Event report message – notify controller of events.
- Logical port statistics request/reply – request and reply of current statistics from a logical port.
- Physical port configuration request – modify the configuration of a physical port.
- Logical port configuration request – modify the configuration of a logical port.
- Physical port capabilities request/reply – request and reply of capabilities of a physical port.
- Drop station – force a mobile station to disassociate.
- Error message – customize error reporting for wireless.

The detailed definitions of the messages are omitted due to space limit.

## C. Implementation

To validate our design and to demonstrate the viability of the ÆtherFlow framework as a platform for the development and deployment of intelligent wireless SDN applications, we implemented and deployed ÆtherFlow on a commercially available access point.

We chose the access point TP-LINK WR1034ND v2 as the hardware platform for our implementation. This AP has five 100Mbps Ethernet ports and one 3-antenna radio interface, supporting protocols IEEE 802.11b/g/n. We replaced the firmware of the AP with OpenWRT 14.07 Barrier Breaker. OpenWRT is an open source Linux distribution designed for network embedded systems. Network utilities are integrated in OpenWRT and are optimized in size to fit in embedded environments which usually do not have as much resources as general purpose computer systems. In OpenWRT, when the radio interface is set up as an access point, its data plane is managed by Linux kernel and its control plane is run in the user space by daemon `hostapd`.

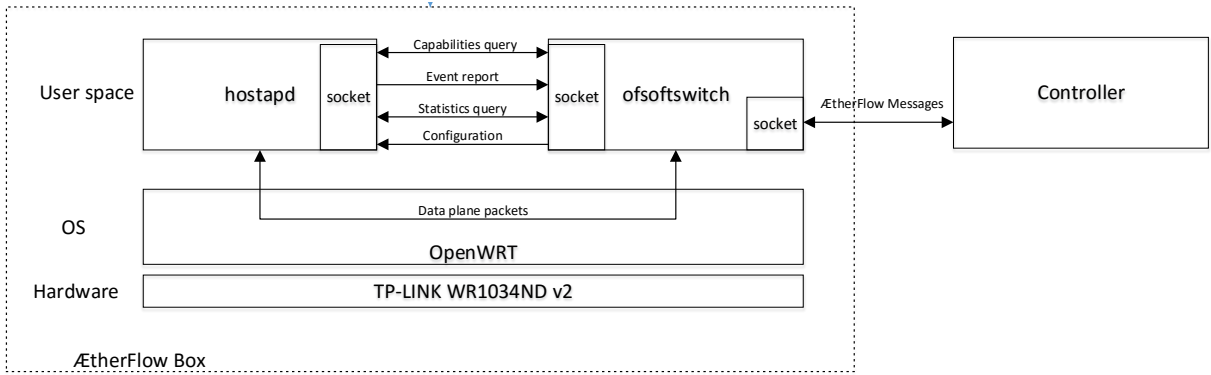

Fig. 3: Implementation of ÆtherFlow.

The native OpenWRT system does not support SDN. To make our access point an SDN switch, we used the open source CPqD SoftSwitch (`ofsoftswitch`), which implements an OpenFlow v1.3 pipeline and switch agent that can be deployed on OpenWRT system.

An ÆtherFlow data plane extension is then implemented in `ofsoftswitch`. The extension adds in wireless physical and logical ports that are mapped to the radio interface and access points managed by `hostapd`. To establish communication between `ofsoftswitch` and `hostapd`, `hostapd` is also modified to enable control of AP from `ofsoftswitch` and event reporting from AP to `ofsoftswitch`. The two processes communicate via a Unix socket in the OpenWRT system. An overview of this implementation is depicted in Figure 3.

Whenever an event related to a mobile station is triggered in `hostapd`, the event summary is sent to `ofsoftswitch`, which forwards it to the controller using the event port message. Whenever a statistics request from the controller is received by `ofsoftswitch`, the request is forwarded to `hostapd`, and the statistics data is sent to `ofsoftswitch` and then sent to the controller with a statistics reply message. Similar behavior occurs for capability queries and configuration updates.

## IV. ÆTHERFLOW APPLICATIONS

The design of ÆtherFlow extends the capability of OpenFlow to wireless (specifically IEEE 802.11) interfaces in a natural way. ÆtherFlow enables applications to control both wireline switches and wireless access points. As a result, network applications that used to require different protocols and cooperation of software from different vendors can now be implemented easily using the ÆtherFlow framework.

We use a Layer 2 fast handoff application to demonstrate the flexibility and new functionality offered by the ÆtherFlow framework. This application aims to facilitate the process of mobile station handoff within the same subnet during which a device's Layer 3 address is not changed.

A typical Layer 2 fast handoff application runs in three phases. The first phase is handoff prediction. The controller collects signal strength information of the mobile stations by requesting statistics of all mobile stations associated with APs under its control. At the same time, it receives the probe signal strength of the mobile stations measured by other APs from the probe event reports. By keeping these data updated in a timely fashion, the controller may predict that a handoff is about to happen, e.g. when the mobile station's signal strength to its associated AP gradually weakens while the signal strength to another AP gradually strengthens.

The second phase of the Layer 2 fast handoff application is multicasting. When a handoff prediction of a mobile client is made, the controller multicasts all the packets with the client as destination to both its current associated AP and the predicted AP. The action is completed by modifying the flow entries of the switches in the network. Multicast guarantees that the client can receive packet immediately after it reassociates with the new AP, thus minimizing the packet loss during the handoff.

The third phase is flow redirection. After the multicasting phase, if the client associates with a new AP, the multicast is stopped and all the following packets to the client will be redirected to the new AP. If the prediction is wrong and a handoff did not occur within a certain timeout period, multicast is stopped and all the following packets will be forwarded to the original AP that the client is associated to. ÆtherFlow makes the decision possible with event report interface that provides client association event report to the controller.

Other than Layer 2 handoff application, wireless network applications such as client steering, user-based QoS control, etc. can also be easily implemented using ÆtherFlow framework.

## V. VALIDATION

We use the ÆtherFlow implementation described in III-C to evaluate the performance and demonstrate the viability of the ÆtherFlow approach. Our results demonstrate that ÆtherFlow framework allows SDN applications to efficiently and dynamically configure wireless networks without loss of performance.

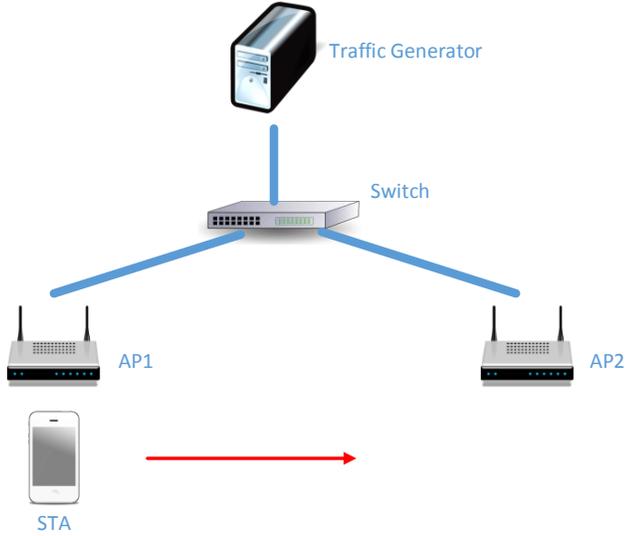

Fig. 4: Experiment network topology

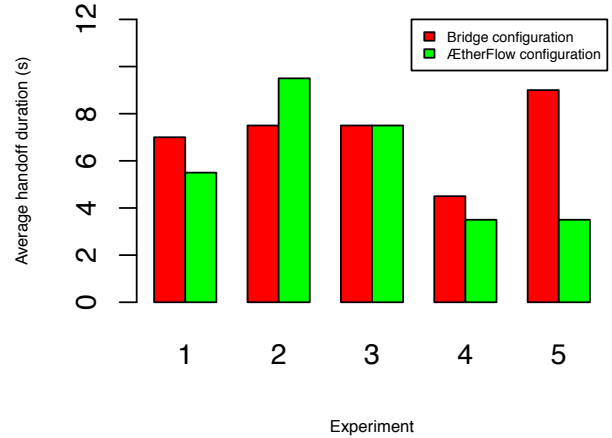

Fig. 5: Comparison of handoff duration in each experiment.

### A. Experiment Setup

Our experiment uses a simple network topology, shown in Figure 4. It consists of two access points (AP1, AP2), a Layer 2 switch, a wireline traffic generator and a wireless 802.11 mobile station (STA). We use an OpenFlow enabled Layer 2 switch and ÆtherFlow enabled APs as described in Section III-C. The mobile station has a single WiFi radio interface.

In our experiment, both APs and the traffic generator are connected to the switch through Ethernet. All the three boxes are connected to an OpenFlow controller through a separate control plane subnet that is not displayed in the figure. The two APs are located at a certain distance and have overlapping coverage areas. Both APs are configured with the same SSID and use open authentication.

### B. Layer 2 Handoff Application

Our Layer 2 handoff application accords with what we described in Section IV. The investigation of good predictors and predictive models for handoff is beyond the scope of this paper. In our implementation, the controller application always predicts that the handoff of STA from AP1 to AP2 will occur seven seconds after the experiment starts. The time period is selected solely for the purpose of this experiment and does not apply for general cases.

After seven seconds, the controller starts to multicast packets going to STA to both AP1 and AP2 by sending FlowMod messages to both APs and the switch. After STA associates with AP2, the controller configures the switch to stop multicasting and forward packets to only AP2. If the predicted handoff did not happen 15 seconds after the prediction, the controller reverts the multicast and forwards packets to only AP1.

### C. Experiment Procedure and Results

In each round of experiment, the mobile station is initially associated with AP1. Both the traffic generator and the mobile station are assigned static IP addresses within the same subnet. Before the experiment starts, a UDP `iperf` session with bandwidth of 9Mbps is initiated from the traffic generator to STA. After experiment starts, STA moves from coverage of AP1 to coverage of AP2, which forces the client to handoff from AP1 to AP2. We move STA in a controlled manner such that the handoff happens at eight seconds after the experiment starts. This time is selected such that the handoff happens one second after the controller application initiates multicasting. Throughput and packet loss rate during each round of test is measured by iperf with an interval of 0.5s. In each round of experiment, one of the following configurations is used:

- **Bridge configuration** uses neither OpenFlow nor ÆtherFlow. Instead, the Layer 2 switch and the two APs use the Linux built-in learning bridge to forward packets. This is the traditional way of configuring a Layer 2 network with two access points and one switch.
- **ÆtherFlow configuration** enables ÆtherFlow on the APs and the switch, and the handoff is managed by the Layer 2 handoff application (described above) running on the ÆtherFlow controller using the Ryu controller framework.

Five rounds of experiments are conducted on each of the two configurations above. In a single round of experiment, the mobile station is considered to be in handoff process during an interval after time $t = 9$s if its average throughput during the interval is less than 8 Mbps. By this criteria we can determine the handoff duration of STA in each round of experiment. Our results, depicted in Figure 5, indicate that the average handoff duration of ÆtherFlow configuration across the five rounds

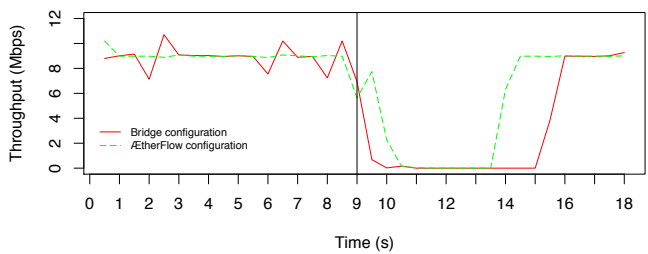

Fig. 6: Comparison of throughput for ÆtherFlow and the baseline configuration.

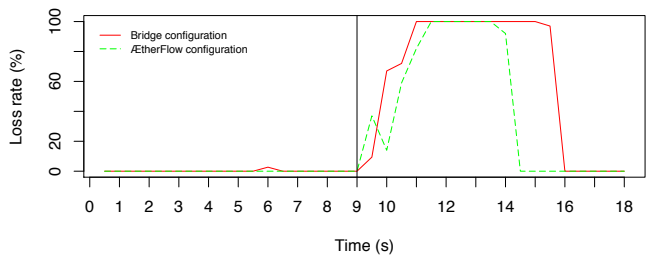

Fig. 7: Comparison of packet loss rate for ÆtherFlow and the baseline configuration.

of experiments is **5.9**s, which is lower than that of bridge configuration **7.1**s.

We compare the traffic throughput and packet loss rate of the two experiments which have median handoff duration in each configuration (experiment 1 for bridge configuration and experiment 5 for ÆtherFlow configuration). The plots are shown in Figures 6 and 7. They demonstrate that in terms of both throughput and loss rate, the ÆtherFlow configuration recovers from handoff faster than the bridge configuration.

The experiment results show that even with the overhead induced by SDN data plane processing, the performance of Layer 2 handoff application based on ÆtherFlow is better that of Linux kernel bridge configuration.

## VI. Conclusion and Future Work

In this paper we presented ÆtherFlow, an SDN framework that provides principled support for wireless networks. ÆtherFlow provides a greater degree of network visibility than traditional SDN approaches. In particular, ÆtherFlow includes the ability to handle wireless packets using an OpenFlow data path, remotely configure access points, query mobile station capabilities and statistics, and report mobile station events.

To validate our ideas, we have implemented an ÆtherFlow switch and adapted an existing OpenFlow controller to work with our extensions of the OpenFlow protocol. We experimented with an SDN-based mobile handoff application, and found that our design slightly outperforms an optimized non-SDN application. We note this is a proof-of-concept experiment designed to show that useful SDN applications can be written against the ÆtherFlow extensions to OpenFlow.

As a general wireless SDN framework, the ÆtherFlow model can also be immediately leveraged to support a number of different applications, or can easily be extended to support them. In addition, similar extension approaches can be used on systems other than IEEE 802.11, such as WiMAX or cellular networks, which is a promising direction for the evolution of SDN. We leave this as our future work.

Our results indicate that while current SDN protocols support the development of very intelligent wireline network management applications, ÆtherFlow is a significant step in bringing that same level of programmability to wireless local area networks.

## References

[1] Cpqd openflow 1.3 software switch. http://cpqd.github.io/ofsoftswitch13/.
[2] C. Jasson Casey, Andrew Sutton, and Alex Sprintson. Tinynbi: Distilling an api from essential openflow abstractions. In *Proceedings of the Third Workshop on Hot Topics in Software Defined Networking*, HotSDN '14, pages 37–42, New York, NY, USA, 2014. ACM.
[3] Open Networking Foundation. Wireless & mobile working group. Technical report.
[4] Open Networking Foundation. Openflow-enabled mobile and wireless networks. Technical report, 2013.
[5] Xin Jin, Li Erran Li, Laurent Vanbever, and Jennifer Rexford. Softcell: Scalable and flexible cellular core network architecture. In *Proceedings of the Ninth ACM Conference on Emerging Networking Experiments and Technologies (CoNEXT '13)*, pages 163–174, 2015.
[6] N. McKeown, T. Anderson, H. Balakrishnan, G. Parulkar, L. Peterson, J. Rexford, S. Shenker, and J. Turner. Openflow: enabling innovation in campus networks. *ACM SIGCOMM Computer Communication Review*, 38(2):69–74, 2008.
[7] Rohan Murty, Jitendra Padhye, Alec Wolman, and Matt Welsh. Dyson: an architecture for extensible wireless lans. In *Proceedings of the 2010 USENIX conference on USENIX annual technical conference (USENIXATC '10)*, pages 15–15, 2010.
[8] Julius Schulz-Zander, Carlos Mayer, Bogdan Ciobotaru, Stefan Schmid, and Anja Feldman. Opensdwn: Programmatic control over home and enterprise wifi. In *1st ACM SIGCOMM Symposium on Software Defined Networking Research (SOSR '15)*, 2015.
[9] Vivek Shrivastava, Nabeel Ahmed, Shravan Rayanchu, Suman Banerjee, Srinivasan Keshav, Konstantina Papagiannaki, and Arunesh Mishra. CENTAUR: Realizing the Full Potential of Centralized WLANs Through a Hybrid Data Path. In *Proceedings of the 15th Annual International Conference on Mobile Computing and Networking (MobiCom '09)*, pages 297–308, 2009.
[10] Lalith Suresh, Julius Schulz-Zander, Ruben Merz, Anja Feldmann, and Teresa Vazao. Towards Programmable Enterprise WLANs with Odin. In *Proceedings of the First Workshop on Hot Topics in Software Defined Networks (HotSDN '12)*, pages 115–120, 2012.
[11] Kok-Kiong Yap, Masayoshi Kobayashi, Rob Sherwood, Nikhil Handigol, Te-Yuan Huang, Michael Chan, and Nick McKeown. Openroads: Empowering research in mobile networks. *ACM SIGCOMM Computer Communication review*, 40(1):125–126, January 2010.
[12] Kok-Kiong Yap, Masayoshi Kobayashi, David Underhill, Srinivasan Seetharaman, Peyman Kazemian, and Nick McKeown. The stanford openroads deployment. In *Proceedings of the 4th ACM International Workshop on Experimental Evaluation and Characterization (WINTECH '09)*, pages 59–66, 2009.
[13] Kok-Kiong Yap, Rob Sherwood, Masayoshi Kobayashi, Te-Yuan Huang, Michael Chan, Nikhil Handigol, Nick McKeown, and Guru Parulkar. Blueprint for introducing innovation into wireless mobile networks. In *Proceedings of the second ACM SIGCOMM workshop on Virtualized Infrastructure Systems and Architectures (VISA '10)*, pages 25–32, 2010.